\documentclass[12pt]{iopart}

\usepackage{graphicx}
\usepackage{bm}
\usepackage{epstopdf}
\usepackage{latexsym}
\usepackage{epsfig}

\newcommand{\ket}[1]{\vert#1\rangle}

\newcommand{\bra}[1]{\left\langle#1\right\vert}

\usepackage{color}
\definecolor{Blue}{rgb}{0,0,1}
\definecolor{Red}{rgb}{1,0,0}
\definecolor{Green}{rgb}{0,1,0}
\definecolor{Purp}{rgb}{.2,0,.2}
\definecolor{white}{rgb}{1,1,1}

\begin{document}
\title{Tomographic characterisation of correlations in a photonic tripartite state}

\author{A. Chiuri$^{1}$, L. Mazzola$^2$, M. Paternostro$^{2,3}$, and P. Mataloni$^{1,4}$}
\address{$^1$ Dipartimento di Fisica, Sapienza Universit\`a di Roma, Piazzale Aldo Moro 5, I-00185 Roma, Italy\\
$^2$ Centre for Theoretical Atomic, Molecular and Optical Physics, School of Mathematics and Physics, Queen's University Belfast, BT7 1NN Belfast, United Kingdom\\
$^3$ Institut f\"ur Theoretische Physik, Albert-Einstein-Allee 11, Universit\"at Ulm, D-89069 Ulm, Germany\\
$^4$ Istituto Nazionale di Ottica (INO-CNR), Largo E. Fermi 6, I-50125 Firenze, Italy}

\begin{abstract}
Starting from a four-partite photonic hyper-entangled Dicke resource, we report the full tomographic characterization of three-, two-, and one-qubit states obtained by 
projecting out part of the computational register. The reduced states thus obtained correspond to fidelities with the expected states larger than $87\%$, therefore 
certifying the faithfulness of the entanglement-sharing structure within the original four-qubit resource. The high quality of the reduced three-qubit state allows for 
the experimental verification of the Koashi-Winter relation for the monogamy of correlations within a tripartite state. 
We show that, by exploiting the symmetries of the three-qubit state obtained upon projection over the four-qubit Dicke resource, such relation can be experimentally 
fully characterized using only 5 measurement settings.  We highlight the limitations of such approach and sketch an experimentally-oriented way to overcome them.
\end{abstract}

\maketitle

\section{Introduction}

Linear optics is at the forefront of current experimental quantum information processing, providing settings, operations and states 
of outstanding quality, controllability and testability~\cite{kok, pan}. Among many other achievements, the first test-bed demonstration of protocols for quantum 
teleportation~\cite{bouwmeester,boschi,riebe}, measurement-based quantum computation~\cite{walther} and quantum-empowered communication~\cite{gao} have been provided 
in linear optics setups. Very recently, some seminal proposals and demonstrations have been given of the suitability of bulk- as well as integrated-optics settings 
for the controllable simulation of quantum phenomena, including the properties of frustrated spin systems~\cite{Ma}, the statistics of two-particle (correlated) 
quantum random walks~\cite{Sansoni}, simple quantum games~\cite{Prevedel} and chemical processes~\cite{Lanyon}. Despite sounding visionary, the possibility of 
using photonic devices to simulate the properties of complex molecular and biological systems, as well the phenomenological features particle statistics and 
high-energy physics is rapidly affirming itself as a viable possibility of not-far-fetched experimental realization~\cite{Aspuru-Guzik,Semiao}.

Such an astonishing success is due to the reconfigurability of photonic interferometers and circuits, as well as the virtually noise-free, highly accurate 
reconstruction of the states that result from the performance of a quantum information protocol via quantum state tomography (QST) techniques~\cite{James}. 
Together with quantum process tomography~\cite{QPT}, which aims at reconstructing the properties of an actual (unitary or non-unitary) quantum channel, QST 
has affirmed itself as the premier diagnostic tool in linear optics quantum information processing, allowing for the reconstruction of the density matrix 
describing the state of intricate multipartite quantum-correlated systems. Besides enabling the evaluation of the performance of most of the schemes mentioned 
above, QST has recently allowed for the exploration of quantum correlations beyond entanglement in two-qubit systems~\cite{Chiuri} and the experimental 
assessment of quantum discord~\cite{Ollivier,Henderson,Dakic} as a resource for quantum communication~\cite{Dakic0}. Further endeavours directed towards the experimental determination of discord in different physical systems have been reported recently~\cite{}. The experimental reconstruction of 
the state of a system, indeed, allows for the handy evaluation of figures of merit that are key to the characterization of the quality of information 
processing at the quantum level. Moreover, it embodies a useful tool for the evaluation of the limitations of a given experimental setup, thus contributing 
to the design and the improvement of the experimental protocol that the set-up is supposed to accommodate.

In this paper, we employ QST to accomplish a twofold task. First, we contribute to the current quest for the experimental generation and manipulation of 
complex multi-qubit entangled states~\cite{Yao, Ceccarelli} by indirectly characterizing the entanglement-sharing structure within a four-qubit photonic 
resource prepared in a symmetric Dicke state with two excitations and encoded in the polarization and momentum degrees of freedom (DOFs) of an hyper-entangled state. 
Four and six-qubit photonic Dicke states have been experimentally generated and tested by means of {\it ad hoc} entanglement witnesses~\cite{Prevedel2,Chiuri2}. 
Very recently, a four-qubit Dicke resource has been used as the computational register for the implementation of important protocols for 
quantum networking~\cite{Chiuri3}. Our approach here will be to characterize the faithfulness of our Dicke-state generator by projecting out a 
growing number of elements of the computational register and testing the faithfulness of the reduced states thus generated to the expected ideal 
states. In particular, we report the experimental QST of three-qubit Dicke states with one and two excitations, as well as two-qubit (Bell) states 
and single-qubit ones. The values off the state fidelity associated with each family of projected states witnesses the closeness of the 
original quadripartite resource to the ideal Dicke one. 

Our second task moves from the high quality of the three-qubit states generated as described above to address the monogamy of  correlations within 
a tripartite state. In particular, we attack the celebrated relation formulated by Koashi and Winter in Ref.~\cite{Koashi2005} and provide its experimental evaluation. 
In particular, we give a state-symmetry simplified expression for each entry of the Koashi-Winter (KW) relation and demonstrate that, for the class of three-qubit 
Dicke states (with one excitation), the latter can be experimentally tested by making use of only five measurements settings, which we have implemented in our 
experiment. We conclude our investigation by providing a plausible analysis of the reasons behind the deviations of the experimental value of the KW relation from 
the expected value. 

The remainder of this paper is organized as follows. In Sec.~\ref{tomo} we introduce the quadripartite Dicke resource, describe the experimental apparatus used 
in order to encode such state in the hyper-entanglement of momentum and polarization of a photonic register and address the projection-based generation and 
tomographic reconstruction of the class of three, two and single-qubit states addressed above.  In Sec.~\ref{KW} we tackle the KW relation and its experimental 
verification over one of our three-qubit states. Besides providing the decomposition of such monogamy relation in terms of the correlators that can be measured 
through only five measurement settings, we highlight the effects of mixedness and anisotropy in its experimental evaluation, providing a KW-based estimate of the 
actual form of the experimentally projected three-qubit resource. Finally, in Sec.~\ref{conclusions}, we draw to our conclusions. 

\section{Experimental realization of a four-qubit Dicke resource and tomographic reconstruction of projected states}
\label{tomo}

\begin{figure}[t]
\centering
\includegraphics[width=\linewidth]{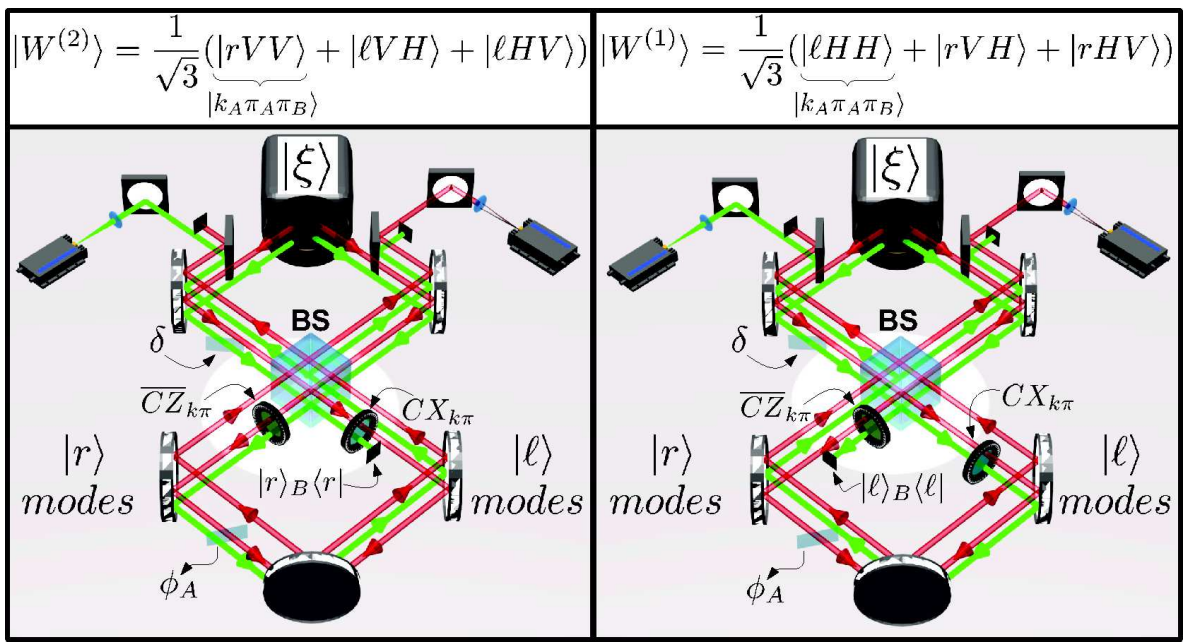}
\caption{Experimental setup for the generation and analysis of three-qubit $\ket{W^{(1,2)}}$ states obtained from the projection of the momentum qubit $d$ of the 
quadripartite Dicke state $\ket{D^{(2)}_4}$ onto $\ket{r}$ (left panel) and $\ket{\ell}$ (right panel), respectively. The red (green) modes represent the optical path
followed by photon A (B).}
\label{setup}
\end{figure}

We used the experimental setup allowing to engineer a symmetric Dicke state by starting from a two photon four qubit polarization-momentum hyper-entangled state ~\cite{Ceccarelli}, already exploited in ~\cite{Chiuri2,Chiuri3,Chiuri4} to test multipartite entanglement, decoherence 
and general quantum correlations~\cite{Chiuri,Chiuri2}. In the following we use both polarization-qubit and path-qubit encoding, $\{\ket{H},\ket{V}\}{\equiv}\{\ket{0},\ket{1}\}$, $\{\ket{r},\ket{\ell}\}{\equiv}\{\ket{0},\ket{1}\}$, with $H/V$ corresponding to the horizontal/vertical polarization of a single photon, while $r$ and $\ell$ are the path followed by photons. The setup was suitably adjusted to prepare the state 
\begin{equation}
\label{enc}
\ket{\xi}_{abcd}{=}[\ket{HH}_{ab}(\ket{r \ell}-\ket{\ell r})_{cd}+2\ket{VV}_{ab}\ket{r \ell}]_{cd}/\sqrt6.
\end{equation} 
Here, qubits $a,c$ ($b,d$) are encoded in the polarization and momentum of photon A (B). Eq.~\eref{enc} can be easily turned into the four-qubit 
two-excitation Dicke state $\ket{D^{(2)}_4}{=}(1/\sqrt6)\sum^6_{j=1}\ket{\Pi_j}$, where $\ket{\Pi_j}$ are the elements of the vector of states built by 
performing all the possible permutations of $0$'s and $1$'s in $\ket{0011}$, by the following unitary transformation~\cite{Chiuri2} 
\begin{equation}
\ket{D^{(2)}_4}{=}{Z}_a\overline{CZ}_{ca}\overline{CZ}_{db}CX_{ca}CX_{db}H_cH_d\ket{\xi}_{abcd}.
\end{equation}
Here $Z_j$ is the $z$-Pauli matrix for qubit $j$, and the controlled-NOT (controlled-PHASE) gate 
$CX_{ij}=\ket{0}_i\bra{0}\otimes I_{j}+\ket{1}_i\bra{1}\otimes X_{j}$ ($\overline{CZ}_{ij}=\ket{1}_i\bra{1}\otimes I_{j}+\ket{0}_i\bra{0}\otimes Z_{j}$) 
is realized by using a half-waveplate with optical axis at $45^{\circ}$ ($0^\circ$) with respect to the vertical direction. The Hadamard gates $H_{c,d}$ 
were implemented by using a polarization insensitive beam-splitter (BS)~\cite{Chiuri3}. In the basis of the physical information carriers, the produced state reads
\begin{equation}
\label{physstate}
\ket{D^{(2)}_4}{=}[\ket{HH\ell\ell}+\ket{VVrr}+(\ket{VH}+\ket{HV})(\ket{r\ell}+\ket{\ell r})]/\sqrt6.
\end{equation}
A full tomographic reconstruction of such state is quite demanding due to the momentum-encoded part of the computational register. However key information 
on its properties and quality can be indirectly gathered by addressing the family of entangled states generated from $\ket{D^{(2)}_4}$ by 
performing projective measurements on part of the qubit register. Let's address this point more in detail. 

It is straightforward to recognize that 
\begin{equation}
\label{DW}
\ket{D^{(2)}_4}=\frac{1}{\sqrt2}\left(\ket{0}_j\ket{W^{(2)}}_{\tilde{j}}+\ket{1}_j\ket{W^{(1)}}_{\tilde{j}}\right),
\end{equation}
with $j$ labeling any of the four qubits in the set $\{a,b,c,d\}$ and $\tilde{j}$ standing for the reduced three-qubit set $\{a,b,c,d\}\slash{j}$. 
Here, $\ket{W^{(2)}}=(\ket{011}+\ket{101}+\ket{110})/\sqrt3$ and $\ket{W^{(1)}}=(\ket{100}+\ket{010}+\ket{001})/\sqrt3$ are three-qubit Dicke states 
with two and one excitation, respectively.  Hence, our strategy will be to test how close are the experimentally generated states by projecting out one 
of the qubits of the computational register to the expected states $\ket{W^{(1)}}$ and $\ket{W^{(2)}}$. On the experimental side we projected out the momentum qubit $d$ by suitably selecting the modes emerging from the BS, as explained in Fig.\ref{setup}. 
\begin{figure}[t]
\centering
\includegraphics[width=0.9\linewidth]{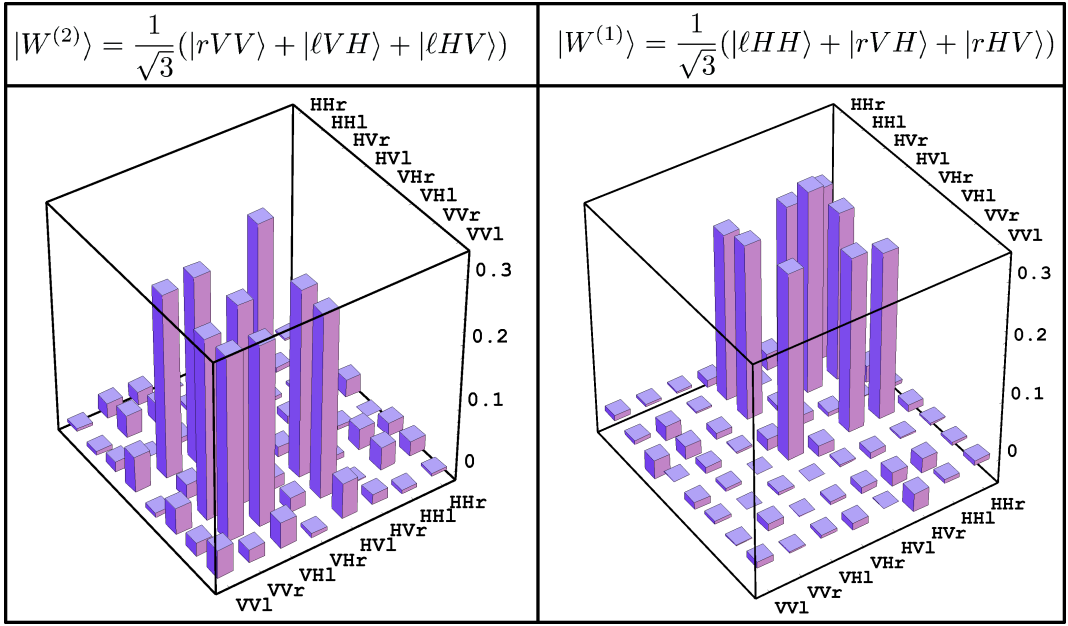}
\caption{Tomographically reconstructed density matrix (real part) of $\ket{W^{(2)}}_{abc}$ (left) and $\ket{W^{(1)}}_{abc}$ (right) obtained by the procedure described in the text. The contributions of the imaginary part are negligible. We have a state fidelity with the target states ${\cal F}=\langle W^{(p)}|\varrho_{p}|W^{(p)}\rangle\simeq0.87$, 
consistently for $p=1,2$.}
\label{TomoW}
\end{figure}

We then performed the tomographic reconstruction of the 
density matrices $\varrho_{1,2}$ describing the states of the remaining three (polarization and momentum) qubits by projecting each of them onto the 
elements of a (statistically complete) subset of 64 states extracted from the 216 ones obtained  
by taking the tensor products of $\{\ket{H},\ket{V},\ket{\pm_x},\ket{\pm_y}\}$ with $\ket{+_k}$ ($\ket{-_k}$) the eigenstates of the $k$-Pauli 
matrix ($k=x,y$) with eigenvalue $+1$ (-1). The two qubits encoded in the polarization DOF were projected onto the necessary elements by using
a usual analysis setup before each detector, composed by a quarter-waveplate, a half-waveplate and a polarizing beam splitter. The
momentum encoded qubit was measured by exploiting the second passage through the BS which allows to project onto the eigenstates
of the necessary Pauli operators. The results of the tomographic analysis are shown in Fig.~\ref{TomoW}, where we also report the 
value of the state fidelity ${\cal F}=\langle W^{(p)}|\varrho_{p}|W^{(p)}\rangle=(0.87\pm 0.01)$, 
consistently for the two projected states. 

\begin{figure}[t]
\centering
\includegraphics[width=0.9\linewidth]{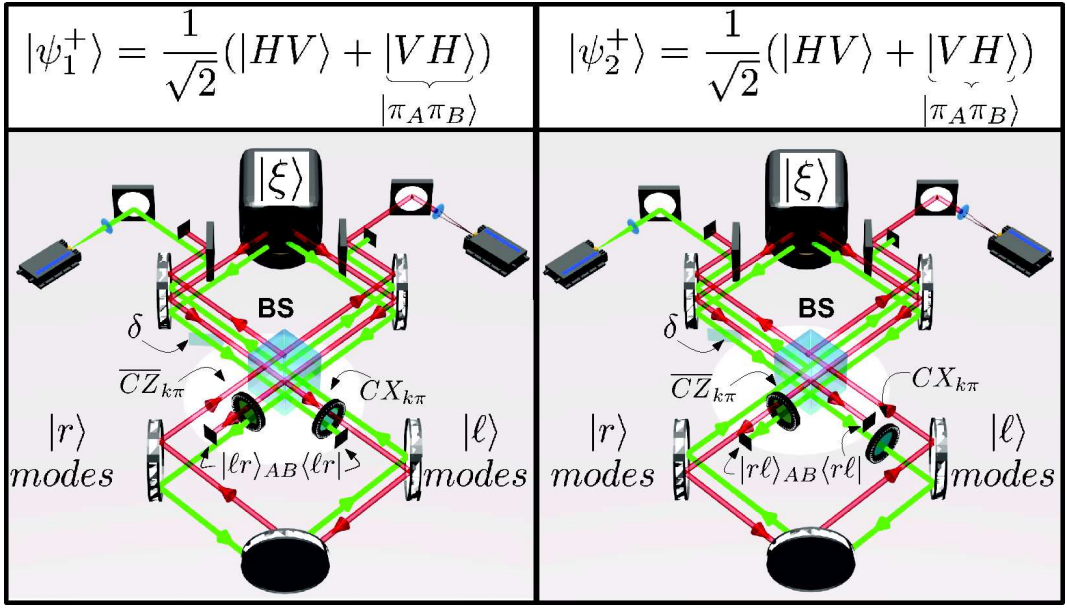}
\caption{Experimental setup used for generation and analysis of the two-qubit $\ket{\psi^+_{1,2}}$ states obtained from the projection of the momentum qubits $c,d$ of the 
quadripartite Dicke state $\ket{D^{(2)}_4}$ onto $\ket{\ell r}_{AB}$ (left) and $\ket{r \ell}_{AB}$ (right), respectively. The red (green) modes represent the 
optical path followed by photon A(B).}
\label{setup2}
\end{figure}

Our study of the inherent entanglement sharing structure within $\ket{D^{(2)}_{4}}$ continues with a further projection. Indeed, by following the same procedure of Eq.~(\ref{DW}), one finds 
\begin{equation}
\ket{D^{(2)}_4}=\frac{1}{\sqrt6}(\ket{0011}_{abcd}+\ket{1100}_{abcd})+\sqrt{\frac{2}{3}}\ket{\psi_+}_{ab}\ket{\psi_{+}}_{cd}
\end{equation}
where we have introduced the Bell state $\ket{\psi^+}{=}(\ket{01}{+}\ket{10})/\sqrt2$. Therefore, by projecting the quadripartite Dicke state 
onto $\ket{01}_{cd}$ or $\ket{10}_{cd}$, we are able to obtain the reduced two-qubit register prepared in $\ket{\psi^+}_{ab}$. We have
experimentally verified that this is indeed the case by projecting the Dicke state created by our state-generation process onto $\ket{10}_{cd}$ 
and $\ket{01}_{cd}$ and  performing a two-qubit QST on the rest of the register. Here the necessary projections were performed 
by suitably selecting the modes emerging from the BS. In the previous case we needed to select only one mode since we projected only qubit $d$. In order to project also qubit $c$ we had to select an optical mode emerging from the BS also for photon A.
\begin{figure}[t]
\centering
\includegraphics[width=0.8\linewidth]{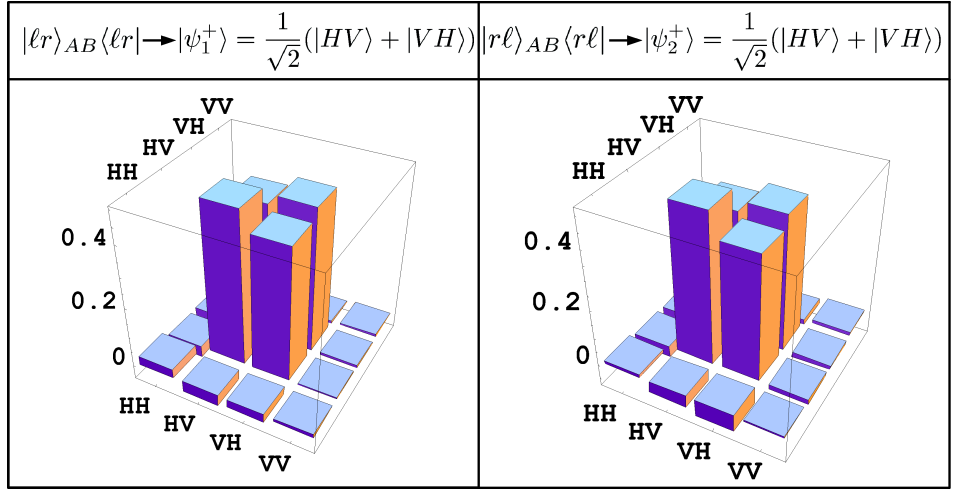}
\caption{Tomographically reconstructed density matrices of $\ket{\psi^+_{1,2}}_{ab}$ achieved by projecting $\ket{D^{(2)}_4}$ onto $\ket{10}_{cd}$ ($\ket{\ell r}_{AB}$, left) 
and $\ket{01}_{cd}$ ($\ket{r \ell}_{AB}$, right), respectively. We show the real part of the elements of each density matrix, the imaginary parts being negligible. The fidelities, 
obtained with respect to the theoretical state, are $\mathcal{F}_{\ket{\psi^+_{1}}} = 0.92 \pm 0.07$ and $\mathcal{F}_{\ket{\psi^+_{2}}}= 0.92 \pm 0.06$. }
\label{TomoBell}
\end{figure}
The results of such analysis are given in Fig.~\ref{TomoBell}, where we have distinguished the reduced states achieved by projecting onto $\ket{10}_{cd}$ or $\ket{01}_{cd}$  by labeling them as $\ket{\psi^+_{1}}$ or $\ket{\psi^+_{2}}$, respectively (needless to say, ideally there should be no difference between them). The associated state fidelity $\ge90\%$, independently of the projection, thus demonstrating the  quality of the reduced two-qubit states. 

\begin{figure}[b]
\centering
\includegraphics[width=0.8\linewidth]{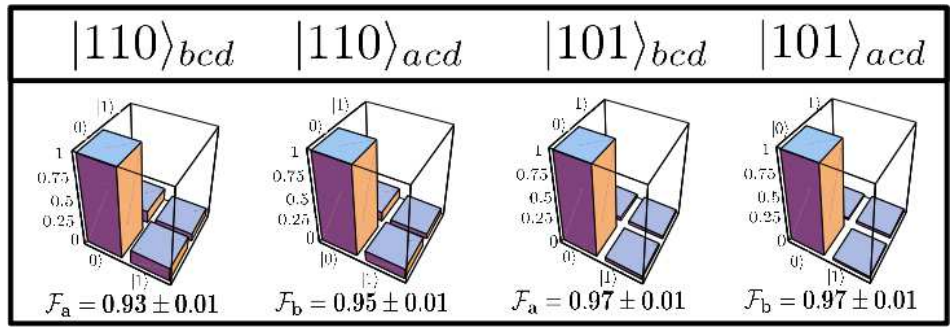}
\caption{Tomographically reconstructed density matrices by projecting three qubits of the Dicke states onto the shown states. 
The fidelities have been calculated with respect to the expected theoretical states. 
We show the real part of the elements of each density matrix, the imaginary parts being negligible.}
\label{Single}
\end{figure}

We conclude this part of our study by describing the results of a further reduction performed on the initial four-qubit Dicke resource. We thus implement 
the projection onto three qubits of the register. We already discussed how the projections onto $\ket{01}_{cd}$ or $\ket{10}_{cd}$ were implemented. 
The last reduction was implemented by selecting the state $\ket{1}_{a(b)}$ and performing the QST on the qubit $b(a)$.
It is worth reminding that qubits $a$ and $b$ were encoded in the polarization DOF, so the projection of these qubits could be performed by selecting 
the physical state $\ket{V}$ before the detector. In Fig.~\ref{Single} we show the results of the projections onto the states 
$\ket{101}_{acd}$, $\ket{110}_{acd}$, $\ket{101}_{bcd}$, $\ket{110}_{bcd}$. Needless to say, the increasing quality of the reduced states is due to the decreasing size of the computational register, which makes state characterisation much more agile, and the filtering effects of the projective measurements~\cite{Prevedel2}. 

\section{Quantum and classical correlations in a tripartite system}

The good quality of the tripartite state produced through the projection above described allows us to go beyond the mere reconstruction of the 
density matrix of the system. In particular, in what follows we concentrate on an analysis of the trade-off between quantum and classical 
correlations in a multi-qubit state, a topic that is currently enjoying a strong and extensive attention by the community interested in 
quantum information processing~\cite{review}.  In the remainder of the paper we thus consider the KW relation~\cite{Koashi2005} and 
apply it to the experimental $|W^{(1)}\rangle$ state. In order to provide a self-contained account of our goal, in what follows we briefly 
introduce the KW relation and identify its context of relevance.


\subsection{The KW relation}

Quantum correlations beyond entanglement have been the centre of an extensive quest for the understanding of their significance, relevance and 
use in the panorama of quantum-empowered information processing and for the study of non-classicality~\cite{review}. Such investigation passes 
through the ab initio definition of quantum correlations \cite{Oppenheim2002,Zurek2003,Groisman2005,Luo2008,Modi2010}, their evaluation for 
different classes of systems~\cite{Luo2008,Bylicka},  and their operational interpretation given in terms of the role that they have 
in quantum communication protocols~\cite{Datta2008,Cavalcanti2011,Piani2011,Mazzola, Chuan}. In this context, in Ref.~\cite{Koashi2005} the 
KW relation was introduced to capture the trade-off between entanglement and classical correlations in a pure tripartite system. This has 
been recently used as the stepping-stone for various results, such as conservation laws and the formulation of a chain-rule for discord~\cite{Fanchini2011,Fanchini2012}. 

The KW relation is based, in some sense, on the monogamous nature of entanglement: if a system $\alpha$ is correlated with both $\beta$ and $\gamma$, 
it cannot be maximally correlated with any of them~\cite{CKW}. 
The KW relation shows that the degree of correlations that, say, $\beta$ can share with the other two is limited by its von Neumann entropy. 
In more quantitative terms and considering only pure tripartite states for the moment, the KW relation reads 
\begin{equation}\label{Koashi-Winter}
J(\beta|\alpha)+E(\beta,\gamma)=S(\beta),
\end{equation}
where any permutation of the systems' indices will lead to similar equalities. Here, $E(\beta,\gamma)$ is the entanglement of formation shared 
by $\beta$ and $\gamma$~\cite{eof}, $S(B)=-\mathrm{Tr}[\rho_\beta \log_2(\rho_\beta)]$ is the von Neumann entropy of $\beta$ only and 
$J(\beta|\alpha)\equiv \max_{{\cal E}_\alpha} J(\beta|\{{\cal E}_\alpha\})$ measures the amount of classical correlations within the state of systems $\alpha$ 
and $\beta$ upon the performance of a measurement (described by the positive-operator-valued-measurement $\{{\cal E}_\alpha\}$) over $\alpha$. 
Here, $J(\beta|\{{\cal E}_\alpha\})=S(\beta)-S(\beta|{\cal E}_\alpha)$, with $S(\beta|{\cal E}_\alpha)$ the quantum conditional entropy of 
the state of system $\beta$~\cite{defin}. The maximization inherent in the definition of $J(\beta|\alpha)$ is necessary to remove any 
dependence on the specific choice of $\{{\cal E}_{\alpha}\}$. 
Moreover, it was shown in \cite{Giorgi} that orthogonal projective measurements are optimal for rank-2 states and provide a very tight bound for rank 3 and 4, therefore the maximisation can be performed within this class of measurements.

As mentioned above, by permuting the systems' labels, we find  five more analogous KW relations. Notice that the relations are six because classical 
correlations are asymmetrical by definition (as one needs to specify the part the measurement is carried on). If the tripartite state is mixed, 
the left-hand side of Eq.~\eref{Koashi-Winter} is always upper-bounded by the right-hand side. In the remainder of this paper we use the quantity defined as the difference between the right and left-hand side of Eq.~\eref{Koashi-Winter}, that is
\begin{equation}\label{KW}
KW=S(\beta)-J(\beta|\alpha)-E(\beta,\gamma).
\end{equation}
Needless to say, the value achieved by a pure state is zero.

\subsection{Experimental investigation of the KW relation}

The maximization necessary for the evaluation of the amount of classical correlations makes the explicit evaluation of the KW relation difficult, in 
general. However, for the particular case of our experimental investigation, state symmetries come to our aids. Indeed, the three-qubit 
reduced states $\varrho_{1,2}$ that we have achieved starting from $\ket{D^{(2)}_4}$ has a large overlap with the ideal $\ket{W^{(1,2)}}$.  
We can thus exploit the symmetries inherent in the latter to derive a manageable expression for the KW relation evaluated over the experiment state. 
Needless to say, we will have the caveat of dealing with the KW inequality rather than Eq.~(\ref{Koashi-Winter}).  In order to fix the ideas, we will consider the subspace of the three-qubit Hilbert space spanned by $\{\ket{001},\ket{010},\ket{100}\}$, so that 
our results will be valid for $\ket{W^{(1)}}$ (although similar conclusions can be drawn for $\ket{W^{(2)}}$). We start evaluating the quantities 
entering Eq.~(\ref{KW}) for a state of systems $\alpha,\beta$ and $\gamma$ of the following form
\begin{equation}
\sigma{=}\left(\begin{array}{cccccccc}
0 & 0 & 0 & 0 & 0 & 0 & 0 & 0 \!\!\!\\
0 & p & c & 0 & c & 0 & 0 & 0 \!\!\!\\
0 & c & p & 0 & c & 0 & 0 & 0 \!\!\!\\
0 & 0 & 0 & 0 & 0 & 0 & 0 & 0 \!\!\!\\
0 & c & c & 0 & p & 0 & 0 & 0 \!\!\!\\
0 & 0 & 0 & 0 & 0 & 0 & 0 & 0 \!\!\!\\
0 & 0 & 0 & 0 & 0 & 0 & 0 & 0 \!\!\!\\
0 & 0 & 0 & 0 & 0 & 0 & 0 & 0 \!\!\!\\
\end{array}\right),
\end{equation}
where we deliberately keep all the quantities symbolic so as to have an expression that depends explicitly on the populations of the state 
$\ket{001}$, $\ket{010}$, and $\ket{100}$ and on the respective coherences. We get
\begin{eqnarray}
\label{components}
S(\beta)&=&-p(2+3 \log_2{p}),\nonumber\\
E(\beta,\gamma)&=&-\frac{1}{2}(1+\sqrt{1-4 p^2})\log_2\left[{\frac{1}{2}(1+\sqrt{1-4
p^2}})\right]\nonumber\\
&-&\frac{1}{2}(1-\sqrt{1-4 p^2})\log_2\left[{\frac{1}{2}(1-\sqrt{1-4
p^2}})\right],\nonumber\\
J(B|A)&=&-p \log_2{p}-2 p \log_2{2 p}\nonumber\\
&+&\frac{1}{2p}\left\{(3 p^2-\sqrt{4 c^2
p^2+p^4})\log_2\left[{\frac{1}{2}\left(1-\frac{\sqrt{4 c^2 p^2+p^4}}{3
p^2}\right)}\right]\right.\nonumber\\
&+&\left.(3 p^2+\sqrt{4 c^2 p^2+p^4})\log_2\left[{\frac{1}{2}\left(1+\frac{\sqrt{4 c^2
p^2+p^4}}{3 p^2}\right)}\right]\right\}.
\end{eqnarray}
The amount of classical correlations were calculated by maximisation over the complete set of orthogonal projectors given by $\Pi_j=\ket{\theta_j}\bra{\theta_j}$ 
with $j=1,2$, $\ket{\theta_1}=\cos{\theta}\ket{0}+e^{\imath \phi}\sin{\theta}\ket{1}$ and $\ket{\theta_2}=e^{-\imath \phi}\sin{\theta}\ket{0}-\cos{\theta}\ket{1}$. 
For the specific state chosen it turns out that the maximisation does not depend on the $\phi$ parameter and is achieved for $\theta=\pi/4$. 
Taking an experimental viewpoint, we notice that the populations $p$ and coherences $c$ can be written explicitly as function of the 
following correlators (calculated over the state of the three-particle system)
\begin{table}[b]
\caption{Table of the correlators needed in order to evaluate the KW relation for symmetric three-qubit states leaving in the 
single-excitation subspace. The experimental values are reported with their associated uncertainties determined by associating 
Poissonian fluctuations to the coincidence counts needed to reconstruct each correlator.}
\vskip0.1cm
\centering
\begin{tabular}{c c}
\hline
\hline
Correlator \qquad & \qquad Value \\
\hline
\hline
$Z_a Z_b Z_c$ \qquad & \qquad $ 0.87 \pm 0.02$
\\
$Z_a Z_b I_c$ \qquad & \qquad $0.35 \pm 0.04$
\\
$Z_a I_b I_c$ \qquad & \qquad $0.26 \pm 0.04$
\\
$X_a X_b Z_c$ \qquad & \qquad $0.55 \pm 0.04$
\\
$Y_a Y_b Z_c$ \qquad & \qquad $0.70 \pm 0.03$
\\
$X_a X_b I_c$ \qquad & \qquad $0.66 \pm 0.03$
\\
$Y_a Y_b I_c$ \qquad & \qquad $0.52 \pm 0.04$
\\
\hline
\hline
\end{tabular}
\label{table}
\end{table}
\begin{eqnarray}
\label{ME}
p=[-\langle Z_\alpha Z_\beta Z_\gamma\rangle+\langle I_\alpha I_\beta I_\gamma\rangle-P(\langle Z_\alpha Z_\beta I_\gamma\rangle)/3+P(\langle Z_\alpha I_\beta I_\gamma\rangle)]/8,\nonumber\\
c=[P(\langle X_\alpha X_\beta Z_\gamma\rangle)+P(\langle Y_\alpha Y_\beta Z_\gamma\rangle)+P(\langle X_\alpha X_\beta I_\gamma\rangle)+P(\langle Y_\alpha Y_\beta I_\gamma\rangle)]/24 \nonumber,
\end{eqnarray}
where $P(\cdot)$ performs the sum of the permutation over $\alpha$, $\beta$ and $\gamma$ of its argument. With this at hand, it is now clear that the KW 
relation can be evaluated by means of only eight correlators. In turn, the latter could be reconstructed experimentally starting from only five measurement settings~\cite{hyllus}. Interestingly, due to the state symmetries, such correlators are exactly those needed to evaluate the fidelity-based 
entanglement witness for $\ket{W^{(1)}}$, when optimally decomposed into local measurement settings~\cite{hyllus}. As such, a single set of local 
measurements would allow for the determination of the genuine tripartite entangled nature of our projected state~\cite{Chiuri3} and the analysis 
of the sharing of correlations among its constituents. In the specific case of our experimental implementation, though, a decomposition into the seven inequivalent settings  that can be identified in Eq.~(\ref{ME}) was more convenient and has thus been used.  

When applied to the experimental state $\varrho_1$, whose experimentally reconstructed correlators are given in Table~\ref{table}, with the 
associations $\alpha=a, \beta=b$ and $\gamma=c$, the combination of the expressions in Eqs.~({\ref{components}) gives us the value 
$KW=0.04\pm0.02$, which is very close to the value expected for $\ket{W^{(1)}}$. Needless to say, this result should be 
considered as a {\it lower bound} to the actual value that Eq.~(\ref{KW}) would take over an experimentally generated state, in 
particular due to the lack of perfect symmetries which played an important role in the determination of the expressions in 
Eqs.~({\ref{components}). This is, again, fully in line with similar issues affecting the evaluation of entanglement witnesses 
over experimental states that do not fulfill the expected degree of symmetry that characterise ideal states. Indeed, the evaluation 
of Eq.~(\ref{KW}) over the tomographically reconstructed state returns a value (averaged over the various permutations of systems' indices) 
of $0.36\pm0.04$, which is significantly different from the value estimated above based on only 8 correlators and the underlying symmetry assumptions. The argument above can be made more quantitative by noticing that, while the numerical functions $J(a|\{\Pi_b\})$ and $J(b|\{\Pi_a\})$ evaluated 
over $\varrho_{1}$ are almost independent of $\phi$ and reach their maximum for $\theta$ close to $\pi/4$, the same does not apply for classical 
correlation involving the $c$ qubit, thus witnessing the asymmetry of the experimental state. This can be ascribed to the fact that qubit 
$c$ is physically encoded in the momentum DOF of a light mode, differently from $a$ and $b$, which are both encoded in polarization.

While we are currently working on a formulation that could encompass possible asymmetries in the tested state yet retaining the 
handiness of an expression that could be experimentally determined by only a handful of measurement settings, we mention that a model that is in 
better agreement with the experimental values of the KW would involve a mixture of $\ket{W^{(1)}}$ with a fully mixed state of three qubits. 
This model is motivated by considering that, experimentally, such projected three-qubit state was created by performing a measurement over a 
quadripartite having a non-ideal estimated fidelity greater than $78\%$ with the ideal Dicke resource and that, for all practical purposes, can be rightly 
written as $\rho_{Dicke}=p(|{D^{(2)}_4}\rangle\langle{D^{(2)}_4}|)+\frac{1-p}{16} I_{16}$~\cite{Chiuri2,Chiuri3}, where the statistical mixture parameter most compatible with the above mentioned fidelity is $p=0.765\pm0.005$. 
It is straightforward to check that such a model provides a value of Eq.~(\ref{KW}) equal to $0.25\pm0.01$, that is significantly closer to the 
actual experimental value, still using only eight correlators overall. By generalising this model to anisotropic noise added to an ideal $\ket{W^{(1)}}$ state, we 
will be able to come up with a broadly applicable expression for the handy experimental estimate of the expectation value of Eq.~(\ref{KW}). Such analysis goes beyond the scopes of this work and will be presented elsewhere~\cite{dopo}.




\section{Conclusions}

We have explored the correlation properties of states obtained by projecting out part of a quadripartite system prepared in a symmetric Dicke state with two excitations and encoded in a high quality photonic hyperentangled state. By designing properly the projections to perform, we have obtained reduced three, two and one-qubit states very close to the expected ones, as shown by the fidelity between the ideal reduced resources and the tomographically reconstructed density matrices of the experimental states. The good quality of the three-qubit Dicke states with one and two excitations obtained in our experiment has paved the way to the experimental verification of a simplified expression for the monogamy of correlations in a tripartite state embodied by the celebrated KW relation~\cite{Koashi2005}. We have highlighted both the advantages, in terms of the very small number of experimental correlators needed for the reconstruction of the expression that we have formulated, and the limitations of our approach, which is heavily based on the symmetries of the ideal state at hand. Our work inspires an experimentally friendly exploration of the correlation properties of multipartite qubit states in state-of-the-art photonic settings characterized by high reconfigurability, exquisite control, and reliable tomographic tools for state diagnostics. 

\label{conclusions}

\section*{Acknowledgments}

This work was supported by EU-Project CHISTERA-QUASAR, PRIN 2009 and FIRB-Futuro in ricerca HYTEQ, the EU under a Marie Curie IEF 
Fellowship, and the UK EPSRC under a Career Acceleration Fellowship and a grant of the ``New Directions for Research Leaders" initiative (EP/G004579/1).


\begin{thebibliography}{99}
\bibitem{eof} For two qubits, the entanglement of formation is given by~\cite{Wootters}
$E\equiv h\left(\frac{1+\sqrt{1-C^2}}{2}\right)$,
with $h(x)=-x\log_2 x-(1-x) \log_2(1-x)$ and $C=\max\{0,\lambda_1-\lambda_2-\lambda_3-\lambda_4\}$ being the concurrence. 
Here, $\{\lambda_i\}$ is the set of eigenvalues (arranged in non-increasing order) of the Hermitian matrix $R=\sqrt{\sqrt{\rho}\tilde{\rho}\sqrt{\rho}}$ 
where $\tilde{\rho}=(\sigma_y\otimes\sigma_y)\rho^{\star}(\sigma_y\otimes\sigma_y)$ and $\rho^{\star}$ is the complex conjugate of the density matrix.
\bibitem{defin} Given $\{{\cal E}_{\alpha\beta}\}$ with elements ${\cal E}^q_\alpha=M_q^{\dagger}M_q$ ($M_p$ being a measurement operator), 
the joint state $\rho_{\alpha\beta}$ is transformed into $\tilde{\rho}_{\alpha\beta}=\sum_q M_q \rho_{\alpha\beta}M_q^{\dagger}$. 
Each measurement outcome occurs with a probability $p_q=\tr({\cal E}^q_\alpha\rho_{\alpha\beta})$ and $\beta$ collapses into the 
post-measurement conditional state $\rho_{\beta|q}=\tr_\alpha({\cal E}^q_\alpha\rho_{\alpha\beta})/p_q$. The quantum conditional 
entropy is thus defined as $S(\beta|{\cal E}_{\alpha})=\sum_qp_qS(\rho_{\beta|q})$.
\end{thebibliography}
\end{document}